# Electric-Field-Switchable Altermagnetism *via* Ligand Rotation in a $d^0$ Metal–Organic Framework

Hongjing Wang,[a] Xiuling Li,[a,*] and Xiaojun Wu[b]

[a] School of Physics and Technology, Nanjing Normal University, Nanjing 210023, China

[b] State Key Laboratory of Precision and Intelligent Chemistry, School of Chemistry and Materials Science, University of Science and Technology of China, Hefei 230026, China

## Abstract

Altermagnetism is an emerging collinear magnetic phase that combines vanishing net magnetization with momentum-dependent spin splitting. While altermagnetic signatures have recently been identified in metal–organic frameworks (MOFs), and a spin-crossover route to switching has been proposed, a non-invasive and reversible external control remains to be demonstrated. Here we report an electric-field-switchable route to altermagnetism in a flexible $d^0$-MOF, where rotation of the organic ligands acts as the switching degree of freedom. In the planar ground state, active $p_z$ orbitals mediate dominant superexchange through the 180° ligand–metal–ligand pathways ($J_2$), stabilizing a conventional antiferromagnetic state. Rotating the ligands out of plane activates the in-plane $p_x/p_y$ orbitals, strongly enhances the 90° pathways ($J_1$), and reverses the $J_1/J_2$ competition. The antiferromagnetic-to-altermagnetic crossover sets in at a moderate tilt angle of ≈ 34°, reaching a $d$-wave spin splitting of 84 meV at the fully rotated 90° configuration. Finite-temperature Monte Carlo simulations establish an ordering scale of $T^* \approx 55$ K for the 90° altermagnetic state, whereas geometric frustration reduces the planar antiferromagnet to decoupled spin chains without finite-temperature long-range order. However, the 90° state lies about 68 meV per ligand above the planar ground state, and stabilizing it therefore requires an external driving force. Ab initio molecular dynamics at room temperature shows that structural thermal fluctuations transiently rotate individual ligands across the threshold but without spatial coherence across the lattice, revealing the dual role of the electric field in stabilizing the metastable state and synchronizing the collective rotation. With polar substituents such as fluorination, this locking is achieved at a critical field of ≈ 0.22 V/Å, comparable to fields applied in field-effect devices and below typical dielectric breakdown limits. Upon field removal, the constrained-rotation energy landscape decreases monotonically back toward 0°, leading to barrierless relaxation to the antiferromagnetic state and completing a reversible switching cycle. Our work establishes a paradigm in which altermagnetic order emerges from ligand *p* orbitals rather than transition-metal *d* orbitals, and structural flexibility enables electric-field control, opening an avenue for responsive organic spintronics.

## 1. Introduction

Altermagnetism has recently been established as a new class of collinear magnetic materials that complements the conventional dichotomy of ferromagnetism and antiferromagnetism.[1-6] Its defining feature is a symmetry-driven arrangement in which opposite spin sublattices are connected by rotational operations, rather than by inversion or translation, enabling nonrelativistic spin splitting in the electronic band structure despite vanishing net magnetization.[1,2] This unique combination endows altermagnets with time-reversal-symmetry-breaking magnetoelectric responses akin to those of ferromagnets, while preserving the advantages of antiferromagnetic order, zero stray fields, robustness against external magnetic perturbations, and terahertz-scale spin dynamics.[7,8] The spin-split bands also allow electrical readout of the magnetic state. These attributes make altermagnets particularly attractive for information storage and spin-based logic devices.[7,8]

Recently, a wide range of altermagnetic systems have been experimentally confirmed or theoretically predicted, spanning inorganic compounds such as rutile-type $RuO_2$, hexagonal semiconductors MnTe and CrSb, and more recently metal-organic frameworks (MOFs) that have attracted considerable attention.[9-15] The vast majority of altermagnetic systems reported to date follow a static design paradigm: the spin splitting is dictated by fixed lattice symmetries and, once synthesized, remains locked in place, producing only a single fixed magnetic state. External-stimulus control has begun to emerge in inorganic and van der Waals platforms, including ferroelectric-field manipulation,[16] twist-angle engineering,[17] bilayer stacking design,[18] strain,[19] and charge-order driving.[20] Nevertheless, a mechanism that enables reversible switching between two distinct magnetic states, such as the 0 and 1 required in memory or logic operations, remains highly desirable, particularly in chemically tunable material platforms.

Metal-organic frameworks (MOFs) are crystalline porous materials constructed from metal nodes and organic linkers through coordination bonds.[21-23] Their chemical modularity offers unique advantages for designing magnetic materials.[24-27] By varying the metal centers,[28,29] ligand types,[30,31] and their connectivity,[32,33] the lattice topology, electronic structure, and magnetic interactions of MOFs can be systematically tuned. More importantly, unlike rigid inorganic crystals, the organic linkers in MOFs are not static structural units. They can undergo conformational changes such as rotation and twisting.[34,35] This inherent structural flexibility provides a physical basis for responding to external stimuli. Recent theoretical work has demonstrated that MOFs can host altermagnetic states.[36-43] Che and co-workers designed organic ligands with nonbonding molecular orbitals to realize d-wave altermagnetic spin splitting in Cr-based 2D MOFs.[36] Baldoví and co-workers exploited ligand symmetry to lower the lattice symmetry, achieving g-wave and *d*-wave altermagnetism in imidazole-based 2D MOFs.[40] These studies confirm that the chemical tunability of MOFs offers a broad design space for altermagnetic materials. However, these existing MOF-AM works still follow the static design logic. They "write" the altermagnetic state into the lattice by selecting specific ligands and metal nodes, leaving the magnetic state fixed once synthesized.[41-43] External control of altermagnetism in MOFs has so far been limited to tuning the spin-polarized currents within a fixed altermagnetic state.[42] Only very recently have two switching strategies been proposed: ferroelectric switching of the altermagnetic spin splitting in a hybrid perovskite MOF, where the control is mediated by the ferroelectric polarization rather than acting directly on the magnetic lattice,[44] and pressure-induced spin-crossover switching in a layered van der Waals MOF, which operates through a spin-state transition localized on the metal centers.[45] Yet a clean, non-invasive, and reversible route to direct electric-field switching between the antiferromagnetic and altermagnetic states remains unexplored.

Here we propose an approach that exploits the conformational flexibility of ligands as a tuning degree of freedom to achieve reversible switching of the altermagnetic state in MOFs. To clearly demonstrate this mechanism, we choose a $d^0$-MOF as our model system. The $d^0$ metal centers carry no magnetic moment, so the magnetic response originates entirely from the p electrons of the ligands. This eliminates the complexity of metal d electrons and allows us to directly track how ligand conformational changes affect the magnetism. In this system, the in-plane (0°) and out-of-plane (90°) configurations of the ligands correspond to different active orbitals, $p_z$ versus $p_x/p_y$. This orbital switching directly alters the relative strength of two superexchange pathways through the metal centers. The 180° pathway ($J_2$) dominates at 0°, stabilizing a conventional antiferromagnetic ground state, while the 90° pathway ($J_1$) is significantly enhanced upon ligand rotation. The $J_1/J_2$ competition reverses at approximately 34°, driving the system into the altermagnetic state. At the fully rotated 90° configuration, the system achieves an 84 meV spin splitting, comparable to the largest ligand-engineered values in MOF altermagnets,[40] yet arising entirely from ligand *p* orbitals rather than metal *d* states. The 90° configuration, however, is energetically higher than the 0° configuration and is metastable. Thermal fluctuations can only transiently populate this state without establishing long-range order. This necessitates an external locking mechanism. By introducing polar substituents on the ligands, we demonstrate that an external electric field of approximately 0.22 V/Å is sufficient to synchronously lock all ligands into the 90° configuration, thereby establishing long-range altermagnetic order. Upon removing the field, the constrained-rotation energy landscape decreases monotonically back to the planar ground state, leading to barrierless relaxation to the 0° antiferromagnetic state and completing a reversible switching cycle. Our work establishes a new paradigm of *p*-electron altermagnetism, advancing the control of altermagnetic states from static materials design to dynamic manipulation of ligand conformation. The combination of MOF structural flexibility, ligand *p*-electron magnetism, and electric-field reversible switching opens a new avenue for organic spintronics.

## 2. Results and discussion

### 2.1 Design concept

The design strategy of our 2D $d^0$-MOF platform is illustrated in **Figure 1**. The framework adopts a square-lattice topology with $d^0$ metal centers coordinated by organic ligands, yielding a metal-to-ligand ratio of 1:2. The metal centers are nonmagnetic, ensuring that all magnetic responses originate solely from the ligand sublattice. This clean electronic structure allows us to isolate the role of ligand geometry in determining the magnetic ground state. **Figure 1a** schematically illustrates the two limiting ligand configurations: the planar configuration (0°) on the left and the perpendicular configuration (90°) on the right, with the ligand $\pi$-orbitals explicitly depicted. In the planar configuration, the ligand $\pi$-orbitals are oriented out of the plane. Two distinct superexchange pathways are identified: the nearest-neighbor pathway $J_1$ (90° ligand-metal-ligand) along the sides of the square, and the next-nearest-neighbor pathway $J_2$ (180° ligand-metal-ligand) along the diagonals. The out-of-plane $\pi$-orbitals are more effectively coupled through the metal centers along the diagonal 180° pathway ($J_2$) than along the sides ($J_1$), making $J_2$ the anticipated dominant exchange channel. This $J_2$-dominated exchange is expected to stabilize an antiferromagnetic (AFM) arrangement, with the magnetic moments on the four ligands adopting an ↑↑↓↓ spin configuration. In the perpendicular configuration, the $\pi$-orbitals reorient to lie in the plane. This reorientation enhances the coupling through the metal centers along the 90° pathway ($J_1$) while suppressing the 180° pathway ($J_2$), reversing the $J_1/J_2$ competition. The resulting $J_1$-dominated exchange is expected to stabilize an altermagnetic (AM) state, with the magnetic moments adopting an alternating ↑↓↑↓ arrangement. The bidirectional arrow between the two configurations indicates the orbital-driven $J_1/J_2$ switching that underlies the magnetic-state transition. **Figures 1b** and **1c** schematically show the corresponding band structures. In the AFM case (Figure 1b), the spin-up and spin-down bands remain degenerate throughout the Brillouin zone. In the AM case (Figure 1c), the degeneracy is lifted, producing momentum-dependent spin splitting, a hallmark of altermagnetism.

**2.2 Structural Model and Magnetic State Analysis**

The 2D MOF monolayer is constructed using Ca centers as the metal nodes and terephthalonitrile (TPN, $C_8N_2H_4$) as the organic linkers. Each TPN ligand coordinates to the Ca centers through its two nitrile nitrogen atoms. The Ca center adopts a square-planar coordination geometry with four N atoms from four distinct TPN ligands, yielding a metal-to-ligand ratio of 1:2. The resulting monolayer features a tetragonal lattice with $C_4$ rotational symmetry and belongs to the $P4$/mmm space group. Owing to the $d^0$ electronic configuration of Ca, the metal centers carry no local magnetic moment. Consequently, all magnetic responses originate entirely from the ligand sublattice, providing a clean platform to investigate the role of ligand geometry in determining the magnetic ground state. The flexible metal-ligand coordination bonds allow the TPN ligands to rotate around the Ca-N bond axis, giving rise to two limiting conformations: the planar configuration (0°, **Figure 2a**), where the ligand molecular plane lies parallel to the *xy* plane, and the perpendicular configuration (90°, **Figure 2d**), where the ligand plane is oriented perpendicular to the *xy* plane. $\sqrt{2}\times\sqrt{2}$ supercell is adopted for both configurations to describe the magnetic structures on an equal footing. The optimized lattice parameters are summarized in **Table S1**. Ab initio molecular dynamics at 300 K confirms that the 0° framework maintains its structural integrity at room temperature (**Figure 2c**). The thermal dynamics of the ligand orientations are analyzed in Section 2.5.

To determine the magnetic ground states of the two limiting configurations, we compared the total energies of four magnetic phases, stripe antiferromagnetic (AFM), altermagnetic (AM), ferromagnetic (FM), and nonmagnetic (NM), for both the 0° and 90° ligand conformations (**Figure S1** and **Table S2**). The AFM state is the lowest in energy for the planar configuration (0°), while the AM state is energetically favored for the perpendicular configuration (90°). Across the two conformations, the planar configuration is energetically favored by 0.27 eV per supercell, identifying the 0° AFM state as the global ground state and the 90° AM state as metastable. Compensating this energetic cost requires an external driving force. As we show in Section 2.6, an electric field acting on appropriately polarized ligands provides such a force.

We then calculated the electronic band structures for both configurations. For the 0° planar configuration (**Figure 2b**), the spin-up and spin-down bands remain fully degenerate throughout the Brillouin zone, consistent with the AFM state. In contrast, the 90° perpendicular configuration (**Figure 2e**) exhibits clear spin splitting along the $M$–$\Gamma$–$M'$ high-symmetry path, with a maximum splitting of ~84 meV at the valence band maximum. This momentum-dependent spin splitting is a hallmark of altermagnetism, confirming the AM nature of the 90° configuration.

The symmetry origin of this splitting can be understood within the spin-group framework, where $[C_2 \| G]$ denotes a $\pi$ rotation in spin space (flipping the spin) combined with a real-space operation $G$ connecting the two magnetic sublattices. In the 0° stripe AFM state, the opposite-spin sublattices are connected by the glide operation $[C_2\|g_x]$ (Figure 2a). Because the centered lattice additionally provides a translation $[C_2\|t]$ between the two sublattices, the bands remain spin degenerate at every $k$ point, consistent with Figure 2b. In the 90° Néel state, no such translation exists and the sublattices are connected instead by the pure fourfold rotation $[C_2\|C_{4z}]$ (Figure 2d), which permits nonrelativistic spin splitting. Notably, the chemical lattice remains fourfold symmetric in both configurations. It is the switch of the magnetic ground state from stripe to Néel order that converts the sublattice connection from translation-type to rotational and thereby activates the spin splitting. The splitting must change sign under $C_{4z}$ and therefore acquires a $d$-wave form factor. The two-dimensional splitting map (**Figure 2f**) resolves nodes along the $k_x$ and $k_y$ axes (through $X$ and $X'$) and maxima along the zone diagonals (toward $M$ and $M'$), identifying the $d_{xy}$ symmetry, consistent with the maximal splitting observed along the $M$–$\Gamma$–$M'$ diagonal path (Figure 2e).

The projected density of states (PDOS) analysis (**Figure 2b** and **2e**, right panels) shows that the electronic states near the Fermi level in both configurations are predominantly contributed by the ligand orbitals, with negligible contribution from the Ca centers. This confirms that the magnetic and electronic properties are governed entirely by the ligand sublattice. The 84 meV splitting is comparable to the largest values achieved in ligand-engineered Cr-based 2D MOF altermagnets (83.9 meV in the $d$-wave and 65 meV in the $g$-wave cases[40]), and its origin from ligand orbitals rather than metal $d$ states highlights the $p$-electron character of the altermagnetism in this $d^0$-MOF platform.

To understand the evolution of the magnetic ground states with ligand rotation, we calculated the relative energies of the AFM, AM, and FM states as a function of the ligand tilt angle (**Figure 3a**). The AFM state remains lowest in energy below ~34°, while the AM state becomes favored above this angle, identifying the critical tilt angle for the AFM-AM transition. To trace the origin of this transition, we extracted the exchange parameters $J_1$ and $J_2$ by mapping the total energies of the magnetic configurations onto a classical Heisenberg model (**Figure 3b**). At 0°, $J_2$ dominates over $J_1$, stabilizing the AFM ground state. As the ligand tilts, $J_1$ increases while $J_2$ is suppressed. The two exchange pathways cross at ~34°, after which $J_1$ dominates and stabilizes the AM state. Quantitatively, ($J_1$, $J_2$) evolves from (5.06, 6.45) meV at 0° to (26.11, −3.94) meV at 90°; notably, $J_2$ changes sign upon rotation, so that the next-nearest-neighbor coupling turns from antiferromagnetic to ferromagnetic and further reinforces the Néel-type AM state. The angular dependence of $J_1$ and $J_2$ thus directly correlates with the magnetic phase transition, confirming the design concept that ligand geometry controls the magnetic ground state through the competition between $J_1$ and $J_2$.

### 2.3 Finite-Temperature Magnetic Ordering

The results presented so far identify the magnetic ground states at zero temperature. Whether the two configurations sustain magnetic order at finite temperature requires a statistical treatment of the spin Hamiltonian. We therefore performed Monte Carlo simulations of the $J_1$–$J_2$ Heisenberg model on the square lattice formed by the ligand moments, with the exchange parameters for the 0° and 90° configurations taken from Figure 3b and converted to classical values as $J_{cl} = J/4$ ($S = 1/2$). The Néel and stripe order parameters and the staggered susceptibility $\chi$ were monitored as functions of temperature. Simulation details are given in the Supporting Information.

For the 90° configuration, the Néel order parameter develops a clear onset with decreasing temperature (**Figure 3c**), and the staggered susceptibility exhibits a pronounced peak at ≈ 55 K (**Figure 3d**), which we identify as the ordering scale $T^*$. Since the model is two-dimensional and isotropic, we refer to $T^*$ as an ordering scale rather than a Néel temperature. Weak magnetic anisotropy and interlayer coupling present in any experimental realization are expected to place the true ordering temperature at or above this scale.

The 0° configuration behaves qualitatively differently. Despite its stripe antiferromagnetic ground state, $m_{\text{stripe}}$ remains small at all temperatures (Figure 3c), and $\chi_{\text{stripe}}$ shows no anomaly down to the lowest temperatures (Figure 3d). The microscopic reason is that on the stripe manifold the $J_1$ contributions cancel exactly by geometric frustration, reducing the low-energy physics to decoupled $S = 1/2$ antiferromagnetic chains running along the 180° exchange pathways, which cannot develop long-range order at any finite temperature. This picture yields an experimentally testable prediction. The uniform magnetic susceptibility of the 0° state should display a broad Bonner-Fisher maximum[46] at $T_{\max} \approx 48$ K, rather than a sharp ordering transition.

The switching demonstrated below therefore toggles not only the zero-temperature ground state but also the finite-temperature magnetic behavior, between a non-ordering state of decoupled spin chains (0°) and a genuinely ordered altermagnet with $T^* \approx 55$ K (90°).

### 2.4 Orbital Mechanism

The microscopic origin of the $J_1/J_2$ reversal is elucidated by examining the spin-resolved charge density and the orbital-resolved density of states for both configurations (**Figure 4**). In the planar configuration (0°, **Figure 4a** and **4b**), the spin charge density is uniformly dispersed over the TPN ligand framework, with spin-polarized electrons predominantly occupying the out-of-plane $p_z$ orbitals of the N and 1,4-site C atoms (**Figure S2**). The electronic states near the Fermi level are dominated by $p_z$ orbitals, while the contributions from in-plane $p_x/p_y$ orbitals are negligible. This out-of-plane orbital distribution enables efficient coupling along the diagonal 180° pathway ($J_2$) through the metal centers, strengthening the $J_2$-mediated exchange and stabilizing the AFM state. In the perpendicular configuration (90°, **Figure 4c** and **4d**), the ligand rotation reorients the $\pi$ orbitals to lie in the plane, exposing the $p_x/p_y$ orbitals that now dominate the electronic states near the Fermi level. This orbital occupation reconstruction enhances the orbital overlap along the 90° pathway ($J_1$) through the metal centers, while simultaneously suppressing the 180° pathway ($J_2$). Consequently, $J_1$ dominates and the AM state is stabilized.

The orbital switching from $p_z$ to $p_x/p_y$ upon ligand rotation thus provides the microscopic mechanism for the $J_1/J_2$ reversal. The spatial reorientation of the ligand $\pi$ orbitals selectively modulates the two superexchange pathways through the metal centers, directly controlling the magnetic ground state.

## 2.5 Thermal Fluctuations

The Monte Carlo analysis above concerns magnetic ordering on fixed ligand configurations. A separate finite-temperature question is whether the ligand orientations themselves remain stable, and whether thermal fluctuations can spontaneously drive the switching. We therefore performed ab initio molecular dynamics (AIMD) simulations at 300 K to assess the structural stability and the thermal behavior of ligand orientations.

As shown in Figure 2c, the total energy trajectory exhibits stable oscillations around a constant value over 10 *ps*, confirming that the 2D framework maintains its structural integrity at room temperature. Despite this stability, thermal fluctuations are sufficient to activate ligand rotation. We sampled 50 snapshots evenly distributed over the trajectory and classified them according to the number of ligands exhibiting large-angle tilts exceeding the critical angle of ~34° (**Figure S3a**). The statistics reveal that single-ligand flip events account for nearly 50% of the sampled snapshots, and dual-ligand flips appear with a similar probability. Three-ligand tilts occur only rarely, while simultaneous four-ligand tilts are not observed within the simulation time scale.

We then extracted the average lifetimes of these thermally activated configurations (**Figure S3b**). Single-ligand flips persist for approximately 600 fs, while dual-ligand flips have an even shorter lifetime of about 400 fs. These values are far too short to establish long-range cooperative order. Moreover, the electronic band structures of the transient configurations (**Figure S4**) show that the spin splitting is substantially reduced compared to the fully ordered 90° AM state (only ~15 meV for the dual-flip case), confirming that thermal fluctuations alone cannot establish a robust AM character.

These results demonstrate a clear limitation. Although thermal fluctuations can transiently drive individual ligands across the switching threshold, the resulting configurations are short-lived, lack spatial coherence, and cannot sustain the long-range orientational order required for the AM state. An external control mechanism is therefore essential to synchronize and stabilize the ligand orientations. This motivates the electric-field approach presented in the following section.

### 2.6 Electric-Field Switching of the Altermagnetic Phase

The preceding sections have shown that ligand rotation can switch the system between AFM and AM, and that thermal fluctuations alone cannot stabilize the AM state because they only produce transient, incoherent local tilts without long-range orientational order. An external control mechanism is therefore required to synchronize and lock the ligand orientations. Electric-field control offers a natural route to achieve this,[47,48] but the pristine TPN ligand is centrosymmetric and lacks a molecular dipole moment, rendering it unresponsive to electric fields. To enable electric-field coupling, we introduce a half-fluorinated TPN ligand, which breaks the inversion symmetry and creates a net dipole moment in the 2D MOF (**Figure 5a**).

We first verified that fluorination does not alter the intrinsic magnetic behavior of the system. The energy evolution, exchange competition, and band structures of the F-system are all consistent with those of the H-system (**Figure 5c** and **Figures S5**-**S7**), confirming that the 0° AFM and 90° AM ground states, as well as the $J_1/J_2$ switching mechanism, remain unchanged upon fluorination. Similar behavior is also observed for methyl substitution (**Figures S8**-**S9**), confirming that the strategy is not limited to fluorination.

We then examined the response of the fluorinated-system to an external electric field applied along the out-of-plane direction (**Figure 5b**). At zero field, the 0° AFM configuration is the ground state, while the 90° AM configuration is metastable. When an electric field is applied, the dipole moment of the fluorinated ligand couples to the field, lowering the energy of the 90° configuration relative to the 0° configuration. We calculated the total energies of the two configurations as a function of field strength (**Figure 5d**). At 0.2 V/Å, the 0° configuration remains lower in energy. The two configurations become nearly degenerate at approximately 0.22 V/Å, beyond which the 90° AM configuration becomes energetically favored, identifying the critical switching field. At 0.3 V/Å, the 90° configuration is clearly stabilized with a significant energy advantage.

To confirm that the altermagnetic character is preserved under the field, we calculated the band structure of the 90° configuration at 0.3 V/Å (**Figure 5e**). The spin splitting along the *M–Γ–M′* path remains around 82 meV (close to the zero-field value), confirming that the electric field locks the ligand orientation without perturbing the AM electronic structure. The finite-temperature magnetic ordering is likewise preserved. Monte Carlo simulations of the fluorinated 90° configuration give an ordering scale of $T^* \approx 54$ K (**Figure 5f**), essentially identical to the H-system value. Under an applied field of 0.3 V/Å, both the staggered susceptibility and the specific heat shift to lower temperature, giving $T^* \approx 47$ K (**Figure S10**), a modest reduction arising from the field-induced modification of the exchange parameters. The switched state nevertheless remains robustly ordered under the field. When the field is removed, the constrained-rotation energy landscape (Figure 3a) decreases monotonically from the 90° configuration toward the planar ground state, so the system relaxes back to the 0° AFM state, establishing full reversibility of the switching.

These results demonstrate that the electric field serves as an effective external switch for the altermagnetic state. It synchronizes the ligand orientations into the 90° configuration and locks them in place, overcoming the thermal disorder that prevents long-range AM order. The switching is reversible, field-tunable, and does not compromise the large AM spin splitting.

We note two limitations of the present study. First, the energy difference between the AFM and AM states at 0° is small (~3 meV), so the magnetic ground-state ordering may be sensitive to the exchange-correlation functional. Benchmarking with hybrid functionals is left for future work. Second, the collective rotation pathway was sampled through constrained relaxations rather than nudged elastic band calculations. The full minimum-energy pathway will be reported elsewhere. The thermal stability discussed here is supported by AIMD at 300 K, consistent with the practice of recent theoretical studies of MOF altermagnets.[14,40]

## Conclusions

We have demonstrated a reversible electric-field switch between antiferromagnetic (AFM) and altermagnetic (AM) states in a $d^0$-metal-organic framework, where the magnetic response originates entirely from the ligand p orbitals. The switching is driven by a purely geometric operation—ligand rotation—which reorients the $\pi$ orbitals from out-of-plane ($p_z$-like) to in-plane ($p_x$/$p_y$-like), thereby reversing the $J_1$/$J_2$ superexchange competition and toggling the magnetic ground state. The resulting AM phase exhibits a large *d*-wave spin splitting of approximately 84 meV, among the largest reported for MOF-based altermagnets and, notably, arising entirely from ligand *p* orbitals. Unlike previous approaches that rely on static design of MOF magnets, our work introduces a dynamic control paradigm for *p*-electron altermagnetism. The ligand rotation is enabled by the structural flexibility inherent to MOFs, and the metastable 90° AM state is locked by a moderate electric field of ~0.22 V/Å. Thermal fluctuations at 300 K can only produce transient local tilts with sub-picosecond lifetimes. The electric field serves as the essential agent to synchronize and lock the ligand orientations into the long-range orientational order required for the AM state. As the constrained-rotation energy landscape decreases monotonically back to the planar AFM ground state, the system is expected to relax back spontaneously upon field removal, completing a reversible switching cycle, while the AM spin splitting remains intact under the field. Monte Carlo simulations further establish an ordering scale of $T^* \approx 55$ K for the 90° altermagnetic state, whereas geometric frustration reduces the planar antiferromagnet to decoupled spin chains that evade finite-temperature long-range order. The electric field therefore switches not only the zero-temperature magnetic ground state but also the finite-temperature magnetic behavior.

By leveraging $d^0$ metal centers, we isolate magnetism to the ligand sublattice, establishing a clean platform for *p*-electron altermagnetism. The observed orbital-selective superexchange switching provides a general mechanism for designing responsive magnetic materials in MOFs, where ligand geometry serves as a direct control knob for magnetic order. More broadly, this strategy can be extended to other polar substituents, highlighting the chemical flexibility of the approach for experimental realization. Our findings open a new avenue for organic spintronics, where the structural flexibility and chemical tunability of MOFs can be exploited for on-demand magnetic switching. The combination of large spin splitting, a moderate switching field, and reversible operation positions this platform as a promising candidate for future low-power, reconfigurable spintronic devices.

**Supporting Information**

Calculation details; torsion-modulated four magnetic orders (Figure S1); C/N sites for PDOS analysis (Figure S2); ligand-flip statistics and lifetimes at 300 K (Figure S3) and band structures of the flipped snapshots (Figure S4); AIMD energy trajectory of the fluorinated $Ca(TPN)_2$ monolayer (Figure S5); band structures, PDOS, and spin charge densities of the fluorinated monolayer (Figures S6 and S7); structural, energetic, and magnetic exchange evolutions of the methylated $Ca(TPN)_2$ monolayer (Figures S8 and S9); finite-temperature magnetic response of the fluorinated 90° configuration under an applied electric field (Figure S10); lattice parameters and magnetic moments (Table S1); relative energies of four magnetic states and $J_1/J_2$ exchange coupling constants (Table S2); lattice, magnetic, and exchange parameters of the methylated monolayer (Table S3) (PDF)

**Corresponding Author**

*Corresponding Email Address: xlli@njnu.edu.cn

**Notes**

The authors declare no competing financial interest.

**Acknowledgment**

This work is supported by the National Natural Science Foundation of China (22225301), the Fundamental Research Funds for the Central Universities (Grants No.WK2490000002). Numerical computations were performed on Hefei advanced computing center.

## References

(1) Šmejkal, L.; Sinova, J.; Jungwirth, T. Beyond Conventional Ferromagnetism and Antiferromagnetism: A Phase with Nonrelativistic Spin and Crystal Rotation Symmetry. *Phys. Rev. X* **2022,** *12*, 031042.

(2) Šmejkal, L.; Sinova, J.; Jungwirth, T. Emerging Research Landscape of Altermagnetism. *Phys. Rev. X* **2022,** *12*, 040501.

(3) Mazin, I. Editorial: Altermagnetism—A New Punch Line of Fundamental Magnetism. *Phys. Rev. X* **2022,** *12*, 040002.

(4) Mazin, I. Altermagnetism Then and Now. *Physics* **2024,** *17*, 4.

(5) Jungwirth, T.; Fernandes, R. M.; Fradkin, E.; MacDonald, A. H.; Sinova, J.; Šmejkal, L. Altermagnetism: An unconventional spin-ordered phase of matter. *Newton* **2025,** *1*, 100162.

(6) Jungwirth, T.; Sinova, J.; Fernandes, R. M.; Liu, Q.; Watanabe, H.; Murakami, S.; Nakatsuji, S.; Smejkal, L. Symmetry, microscopy and spectroscopy signatures of altermagnetism. *Nature* **2026,** *649*, 837-847.

(7) Bai, L.; Feng, W.; Liu, S.; Šmejkal, L.; Mokrousov, Y.; Yao, Y. Altermagnetism: Exploring New Frontiers in Magnetism and Spintronics. *Adv. Funct. Mater.* **2024,** *34*, 2409327.

(8) Song, C.; Bai, H.; Zhou, Z.; Han, L.; Reichlova, H.; Dil, J. H.; Liu, J.; Chen, X.; Pan, F. Altermagnets as a new class of functional materials. *Nat. Rev. Mater.* **2025,** *10*, 473-485.

(9) Fedchenko, O.; Minár, J.; Akashdeep, A.; D'Souza, S. W.; Vasilyev, D.; Tkach, O.; Odenbreit, L.; Nguyen, Q.; Kutnyakhov, D.; et al. Observation of time-reversal symmetry breaking in the band structure of altermagnetic $RuO_2$. *Sci. Adv.* **2024,** *10*, eadj4883.

(10) Krempasky, J.; Smejkal, L.; D'Souza, S. W.; Hajlaoui, M.; Springholz, G.; Uhlirova, K.; Alarab, F.; Constantinou, P. C.; Strocov, V.; et al. Altermagnetic lifting of Kramers spin degeneracy. *Nature* **2024,** *626*, 517-522.

(11) Milivojević, M.; Orozović, M.; Picozzi, S.; Gmitra, M.; Stavrić, S. Interplay of altermagnetism and weak ferromagnetism in two-dimensional $RuF_4$. *2D Mater.* **2024,** *11*, 035025.

(12) Reimers, S.; Odenbreit, L.; Smejkal, L.; Strocov, V. N.; Constantinou, P.; Hellenes, A. B.; Jaeschke Ubiergo, R.; Campos, W. H.; Bharadwaj, V. K.; et al. Direct observation of altermagnetic band splitting in CrSb thin films. *Nat. Commun.* **2024,** *15*, 2116.

(13) Gao, Z. F.; Qu, S.; Zeng, B.; Liu, Y.; Wen, J. R.; Sun, H.; Guo, P. J.; Lu, Z. Y. AI-accelerated discovery of altermagnetic materials. *Natl. Sci. Rev.* **2025,** *12*, nwaf066.

(14) Lopez-Alcala, D.; Shumilin, A.; Baldovi, J. J. Altermagnetic Metal-Organic

Frameworks. *J. Am. Chem. Soc.* **2026,** *148*, 27008-27022.
(15) Ni, X.; Ji, H.; Liu, F.; Bredas, J. L. Emergence of g-Wave Altermagnetism in Three-Dimensional Metal-Organic Frameworks. *J. Am. Chem. Soc.* **2026,** *148*, 15417-15425.
(16) Zhu, Z.; Duan, X.; Zhang, J.; Hao, B.; Zutic, I.; Zhou, T. Two-Dimensional Ferroelectric Altermagnets: From Model to Material Realization. *Nano Lett.* **2025,** *25*, 9456-9462.
(17) Liu, Y.; Yu, J.; Liu, C. C. Twisted Magnetic Van der Waals Bilayers: An Ideal Platform for Altermagnetism. *Phys. Rev. Lett.* **2024,** *133*, 206702.
(18) Zeng, S.; Zhao, Y. Bilayer stacking A-type altermagnet: A general approach to generating two-dimensional altermagnetism. *Phys. Rev. B* **2024,** *110*, 174410.
(19) Zhang, W.; Zheng, M.; Liu, Y.; Zhang, Z.; Xiong, R.; Lu, Z. Strain-induced nonrelativistic altermagnetic spin splitting effect. *Phys. Rev. B* **2025,** *112*, 024415.
(20) Dey, D.; Choudhary, S.; Ramasesha, S.; Raghunathan, R. Charge-order-driven altermagnetism in a bipartite lattice. *Phys. Rev. B* **2025,** *111*, 214439.
(21) Thorarinsdottir, A. E.; Harris, T. D. Metal-Organic Framework Magnets. *Chem. Rev.* **2020,** *120*, 8716-8789.
(22) Song, X.; Liu, J.; Zhang, T.; Chen, L. 2D conductive metal-organic frameworks for electronics and spintronics. *Sci. China Chem.* **2020,** *63*, 1391-1401.
(23) Chakraborty, G.; Park, I. H.; Medishetty, R.; Vittal, J. J. Two-Dimensional Metal-Organic Framework Materials: Synthesis, Structures, Properties and Applications. *Chem. Rev.* **2021,** *121*, 3751-3891.
(24) Lopez-Cabrelles, J.; Manas-Valero, S.; Vitorica-Yrezabal, I. J.; Siskins, M.; Lee, M.; Steeneken, P. G.; van der Zant, H. S. J.; Minguez Espallargas, G.; Coronado, E. Chemical Design and Magnetic Ordering in Thin Layers of 2D Metal-Organic Frameworks (MOFs). *J. Am. Chem. Soc.* **2021,** *143*, 18502-18510.
(25) Coronado, E. Molecular magnetism: from chemical design to spin control in molecules, materials and devices. *Nat. Rev. Mater.* **2020,** *5*, 87-104.
(26) Wang, M.; Dong, R.; Feng, X. Two-dimensional conjugated metal-organic frameworks (2D c-MOFs): chemistry and function for MOFtronics. *Chem. Soc. Rev.* **2021,** *50*, 2764-2793.
(27) Che, Y.; Lv, H.; Wu, X. Precision Chemical Design for Reticular Spintronics. *Precis. Chem.* **2026**, DOI: 10.1021/prechem.6c00044.
(28) Li, W.; Sun, L.; Qi, J.; Jarillo-Herrero, P.; Dinca, M.; Li, J. High temperature ferromagnetism in π-conjugated two-dimensional metal-organic frameworks. *Chem. Sci.* **2017,** *8*, 2859-2867.
(29) Li, X.; Li, X.; Yang, J. Two-Dimensional Multifunctional Metal-Organic Frameworks with Simultaneous Ferro-/Ferrimagnetism and Vertical Ferroelectricity. *J. Phys. Chem. Lett.* **2020,** *11*, 4193-4197.
(30) Lv, H.; Wu, D.; Cui, X.; Wu, X.; Yang, J. Enhancing Magnetic Ordering in Two-Dimensional Metal-Organic Frameworks via Frontier Molecular Orbital Engineering. *J. Phys. Chem. Lett.* **2024,** *15*, 9960-9967.
(31) Lv, H.; Li, X.; Wu, D.; Liu, Y.; Li, X.; Wu, X.; Yang, J. Enhanced Curie Temperature of Two-Dimensional Cr(II) Aromatic Heterocyclic Metal-Organic Framework Magnets via Strengthened Orbital Hybridization. *Nano Lett.* **2022,** *22*,

1573-1579.
(32) Gu, L.; Xiong, Z.; Zhong, Y.; Chen, Z. Reticular Synthesis of High-Connectivity Metal-Organic Frameworks with Kuratowski-Type Building Blocks. *J. Am. Chem. Soc.* **2026,** *148*, 99-105.
(33) Kong, X. J.; Xie, H.; Liu, J.; He, T.; Wang, X.; Wang, K.; Tang, X.; Hou, B.; Kirlikovali, K. O.; et al. Torsional Flexibility Tuning of Hexa-Carboxylate Ligands to Unlock Distinct Topological Access to Zirconium Metal-Organic Frameworks. *J. Am. Chem. Soc.* **2026,** *148*, 3562-3569.
(34) Klokic, S.; Naumenko, D.; Marmiroli, B.; Carraro, F.; Linares-Moreau, M.; Zilio, S. D.; Birarda, G.; Kargl, R.; Falcaro, P.; et al. Unraveling the timescale of the structural photo-response within oriented metal-organic framework films. *Chem. Sci.* **2022,** *13*, 11869-11877.
(35) Mahato, B.; Dinda, S.; Maiti, A.; Kumar, R.; Ghoshal, D. Amine-Imine Tautomeric Excited State Intramolecular Proton Transfer in Metal-Organic Frameworks: Alcohol and Anion Recognition. *Chem. Eur. J.* **2025,** *31*, e202404141.
(36) Che, Y.; Chen, Y.; Liu, X.; Lv, H.; Wu, X.; Yang, J. Inverse Design of 2D Altermagnetic Metal-Organic Framework Monolayers from Huckel Theory of Nonbonding Molecular Orbitals. *JACS Au* **2025,** *5*, 381-387.
(37) Che, Y.; Lv, H.; Wu, X.; Yang, J. Bilayer Metal-Organic Framework Altermagnets with Electrically Tunable Spin-Split Valleys. *J. Am. Chem. Soc.* **2025,** *147*, 14806-14814.
(38) Che, Y.; Lv, H.; Wu, X.; Yang, J. Engineering Altermagnetic States in Two-Dimensional Square Tessellations. *Phys. Rev. Lett.* **2025,** *135*, 036701.
(39) Che, Y.; Guo, Y.; Lv, H.; Wu, X.; Yang, J. Symmetry-Driven Multiferroic Altermagnetism in Two-Dimensional Materials. *J. Am. Chem. Soc.* **2026,** *148*, 5125-5131.
(40) Lopez-Alcala, D.; Ruiz, A. M.; Shumilin, A.; Baldovi, J. J. Chemical Engineering of Altermagnetism in Two-Dimensional Metal-Organic Frameworks. *J. Am. Chem. Soc.* **2026,** *148*, 25415–25425.
(41) Xu, P.; Che, Y.; Lv, H.; Wu, X.; Yang, J. Engineering Altermagnetic Transitions in Two-Dimensional Metal-Organic Frameworks via Chemical Symmetry Breaking. *J. Am. Chem. Soc.* **2026,** *148*, 30675-30681.
(42) Che, Y.; Lv, H.; Wu, X.; Yang, J. Realizing altermagnetism in two-dimensional metal-organic framework semiconductors with electric-field-controlled anisotropic spin current. *Chem. Sci.* **2024,** *15*, 13853-13863.
(43) Zhang, Z.; Sun, H.; Liu, K.; Zhang, L.; Dong, M.; Wang, A.; Shao, X.; Zhao, M. Lifshitz transition and anisotropic plasmons in altermagnetic two-dimensional Cairo pentagonal metal-organic frameworks. *Phys. Rev. B* **2026,** *113*, 035411.
(44) Gu, M.; Liu, Y.; Zhu, H.; Yananose, K.; Chen, X.; Hu, Y.; Stroppa, A.; Liu, Q. Ferroelectric Switchable Altermagnetism. *Phys. Rev. Lett.* **2025,** *134*, 106802.
(45) López-Alcalá, D.; Ruiz, A. M.; Shumilin, A.; Baldov í , J. J. Switchable Altermagnetism in a Layered van der Waals Metal-Organic Framework Driven by Spin-Crossover. **2026**, arXiv：2609.07548. https://doi.org/10.48550/arXiv.2609.07548
(46) Bonner, J. C.; Fisher, M. E. Linear Magnetic Chains with Anisotropic Coupling.

*Phys. Rev.* **1964,** *135*, A640-A658.
(47) Durholt, J. P.; Jahromi, B. F.; Schmid, R. Tuning the Electric Field Response of MOFs by Rotatable Dipolar Linkers. *ACS Cent. Sci.* **2019,** *5*, 1440-1448.
(48) Jutglar-Lozano, K.; Deumal, M.; Ribas-Arino, J.; Bromley, S. T. Rational Design of Electric Field-Responsive Building Blocks for All-Organic 2D Magnetoelectric Materials. *J. Am. Chem. Soc.* **2025,** *147*, 22550-22561.

# Figures

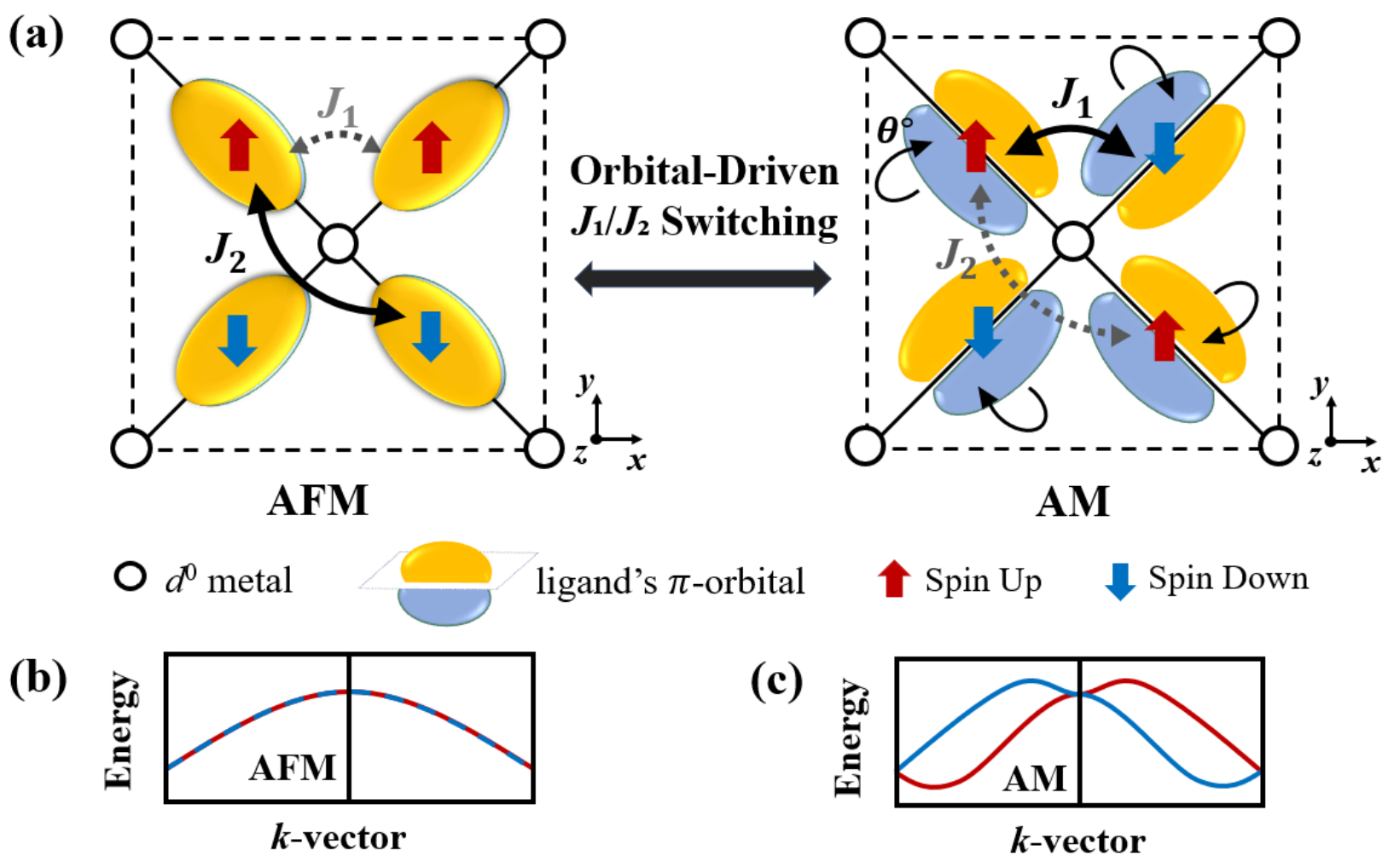

**Figure 1.** Design concept of the 2D $d^0$-MOF. (a) Schematic of the two limiting ligand configurations: the planar configuration (0°, left) with AFM spin ordering and the perpendicular configuration (90°, right) with AM spin ordering. The nearest-neighbor $J_1$ (90° ligand-metal-ligand) and next-nearest-neighbor $J_2$ (180° ligand-metal-ligand) pathways are marked; bold black lines highlight the dominant exchange channel in each case ($J_2$ for 0°, $J_1$ for 90°). (b) Schematic band structure of the AFM state showing spin-degenerate bands. (c) Schematic band structure of the AM state showing momentum-dependent spin splitting.

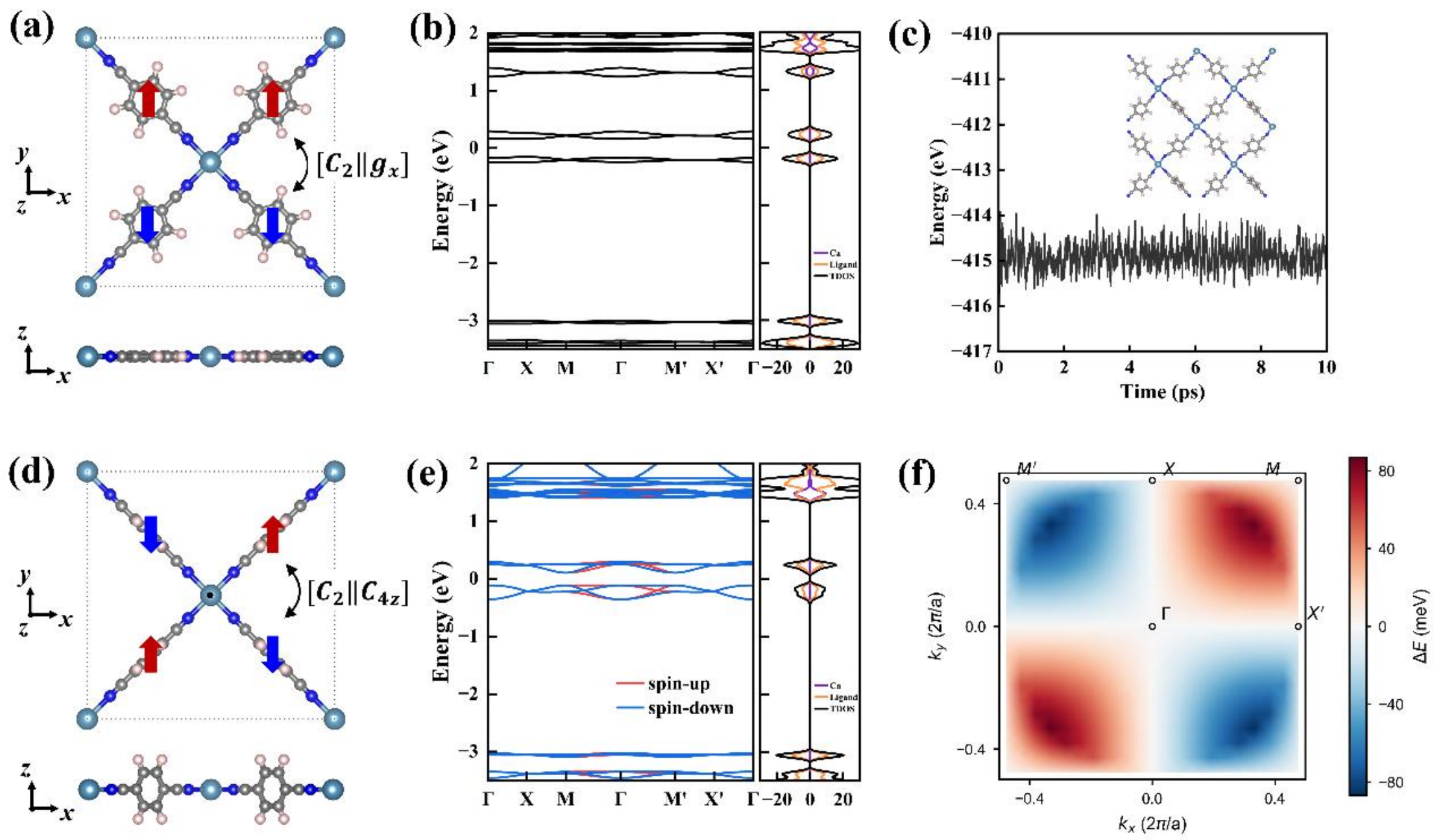

**Figure 2.** Structure and electronic properties of the $Ca(TPN)_2$ monolayer. (a) Top and side views of the planar configuration (0°) in the AFM ground state. Teal, blue, gray, and white spheres represent Ca, N, C, and H atoms, respectively, and red and blue arrows denote spin-up and spin-down moments on the ligands. A √2×√2 supercell is adopted for both configurations, and the connecting spin-symmetry operations are indicated. (b) Band structure and PDOS of the 0° configuration. (c) Total-energy trajectory from AIMD simulations at 300 K. The inset shows the structural snapshot after 10 *ps*. (d) Top and side views of the perpendicular configuration (90°) in the AM ground state, using the same color code as in (a). (e) Band structure and PDOS of the 90° configuration, with spin-up and spin-down bands shown in red and blue, respectively. (f) Two-dimensional spin-splitting map $\Delta E(k)$ of the top valence band over the Brillouin zone for the 90° configuration.

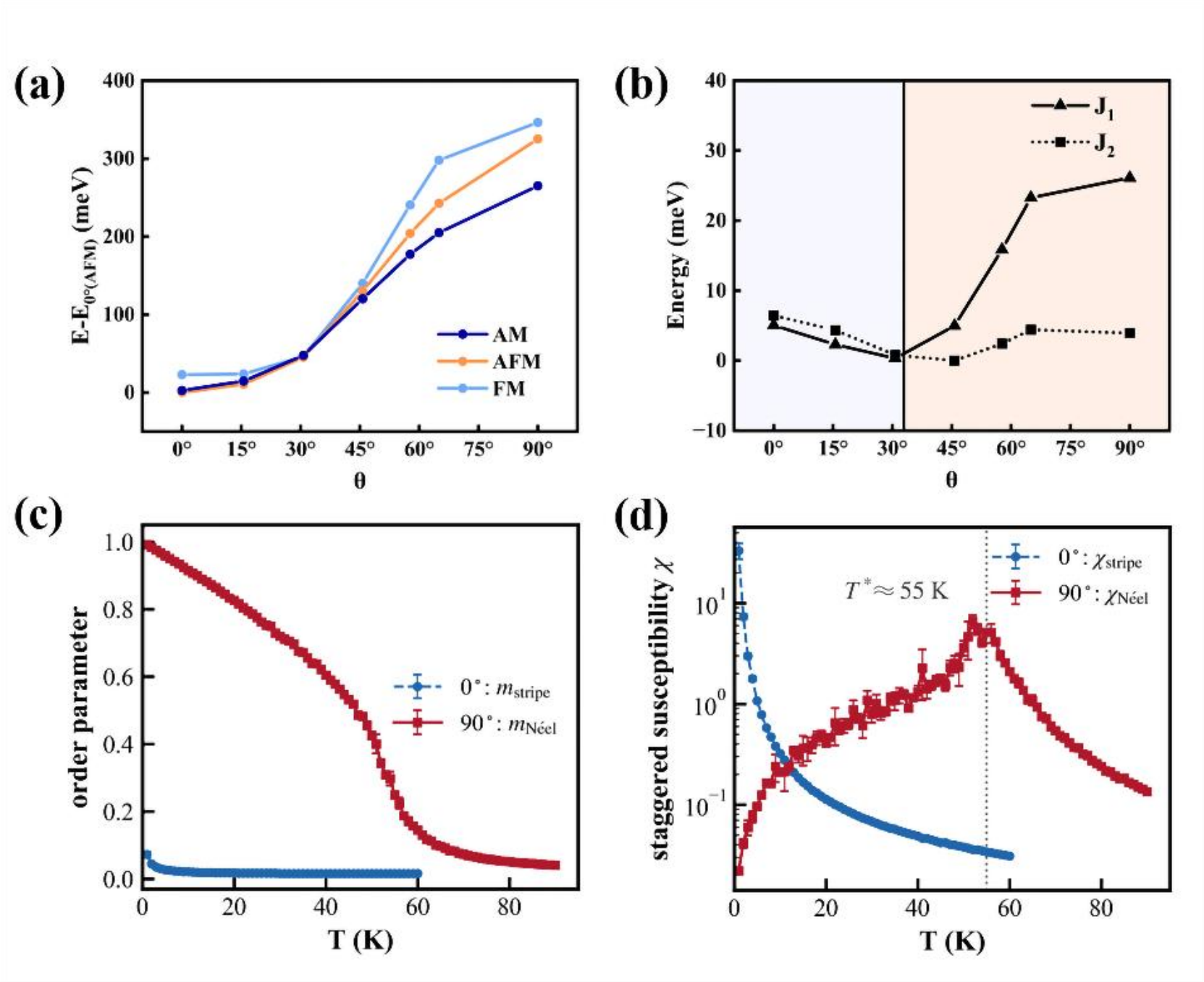

**Figure 3.** Magnetic ground-state evolution and finite-temperature ordering of the Ca(TPN)$_2$ monolayer. (a) Relative energies of the AFM, AM, and FM states as a function of the ligand tilt angle θ, with the 0° AFM state set as zero. (b) Exchange parameters $J_1$ and $J_2$ as a function of θ. The blue- and orange-shaded regions denote the AFM and AM ground-state regimes, respectively. (c) Temperature dependence of the Monte Carlo order parameters $m_{\mathrm{stripe}}$ (0°, blue) and $m_{\mathrm{Néel}}$ (90°, red). (d) Staggered susceptibilities $\chi_{\mathrm{stripe}}$ (0°) and $\chi_{\mathrm{Néel}}$ as a function of temperature, using the same color code as in (c). The vertical dotted line marks the ordering scale $T^* \approx 55$ K of the 90° configuration, while the 0° configuration shows no signature of finite-temperature long-range order.

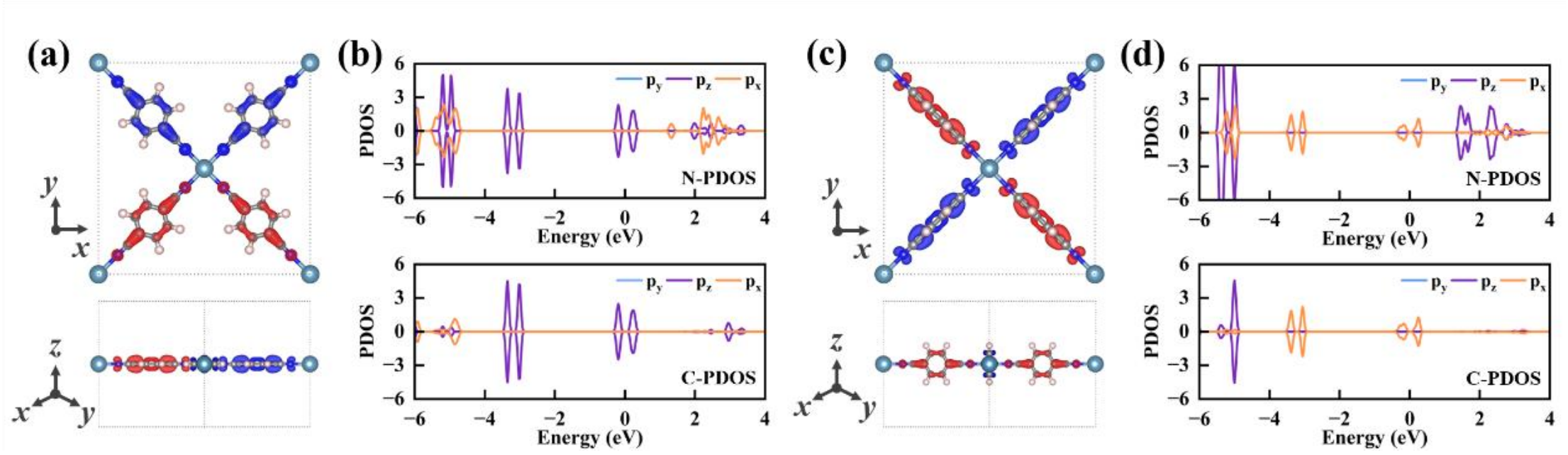

**Figure 4.** Spin charge density and projected density of states (PDOS) of the $Ca(TPN)_2$ monolayer. (a,c) Top and side views of spin charge density for the 0° and 90° configurations, respectively. Red and blue isosurfaces denote spin-up and spin-down charge densities (isovalue = 0.004 e/bohr$^3$). (b,d) PDOS of the N and C atoms that contribute most to the magnetic moment in the 0° and 90° configurations, respectively.

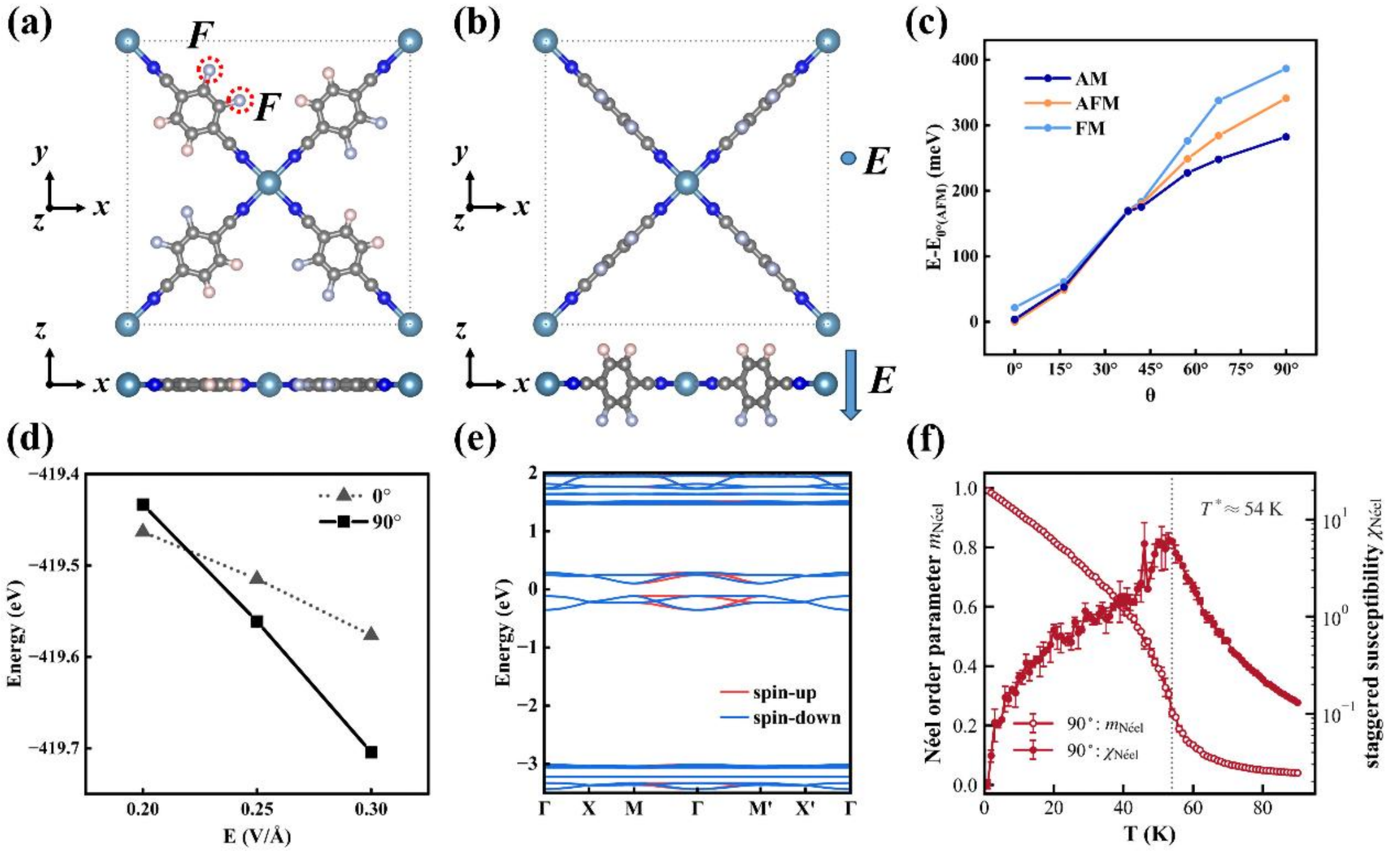

**Figure 5.** Electric-field switching in the fluorinated Ca(TPN)$_2$ monolayer. (a) Top and side views of the fluorinated monolayer in the planar (0°) configuration, with the F atoms highlighted. (b) Top and side views of the 90° configuration under an out-of-plane electric field. The blue-gray arrow indicates the direction of the applied field. (c) Relative energies of the AFM, AM, and FM states as a function of the ligand tilt angle θ, with the 0° AFM state set as zero. (d) Total energies of the 0° and 90° configurations as a function of the applied electric field. (e) Band structure of the 90° configuration under a field of 0.3 V/Å, with spin-up and spin-down bands shown in red and blue, respectively. (f) Néel order parameter $m_{\text{Néel}}$ (left axis) and staggered susceptibility $\chi_{\text{Néel}}$ (right axis) of the 90° configuration as a function of temperature at zero field.

# Supporting Information

## Electric-Field-Switchable Altermagnetism *via* Ligand Rotation in a $d^0$ Metal–Organic Framework

Hongjing Wang,[a] Xiuling Li,[a,*] and Xiaojun Wu[b]

[a] School of Physics and Technology, Nanjing Normal University, Nanjing 210023, China

[b] State Key Laboratory of Precision and Intelligent Chemistry, School of Chemistry and Materials Science, University of Science and Technology of China, Hefei 230026, China

## Table of Contents

## Calculation details

First-principles density functional theory (DFT) calculations were performed using the Vienna Ab initio Simulation Package (VASP).[1,2] The exchange-correlation interactions were described by the Perdew–Burke–Ernzerhof (PBE) generalized gradient approximation[3], while Grimme's DFT-D3 dispersion correction was included to capture van der Waals forces[4]. A vacuum region of ~15 Å along the z-axis was introduced to suppress spurious interlayer coupling between periodic images. The plane-wave cutoff energy was fixed at 500 eV. Structural relaxation proceeded until the total energy converged to $10^{-5}$ eV and the residual atomic forces fell below 0.01 eV/Å. A Monkhorst–Pack k-point mesh of 3×3×1 was used for both geometry optimization and electronic structure calculations.[5] Ab initio molecular dynamics (AIMD) simulations in the canonical (NVT) ensemble at 300 K were implemented to evaluate the thermal stability of the monolayer. Temperature regulation was realized via the Nosé–Hoover thermostat[6-9], and the same 3×3×1 k-grid was adopted for dynamic simulations. All output files were post-processed with the VASPKIT.[10]

Monte Carlo simulations of the $J_1$–$J_2$ Heisenberg model were performed on $L \times L$ square lattices ($L$ = 40 and 60) with periodic boundary conditions. The quantum exchange parameters were converted to classical values as $J_{cl} = J/4$, corresponding to $S$ = 1/2 ligand moments treated as classical unit vectors. Trial moves rotate individual spins within a cone around the current spin direction, with an adaptive half-angle (initial value 0.3 rad) that keeps the acceptance ratio close to 50%. One Monte Carlo sweep updates every spin once via a checkerboard decomposition. At each temperature, $2\times10^4$ sweeps were discarded for thermalization, followed by $3\times10^4$ sweeps for measurement. Three independent runs with different random seeds were averaged, except for the field-on calculation, which used a single run. The Néel and stripe order parameters $m_{\text{Néel}}$ and $m_{\text{stripe}}$ were monitored, together with the staggered susceptibility $\chi = N(\langle m^2\rangle - \langle|m|\rangle^2)/k_BT$ and the specific heat $C_V = N(\langle e^2\rangle - \langle e\rangle^2)/k_BT^2$ ($e$ = energy per spin), both reported per spin.

Given that all magnetic interactions are exclusively mediated by superexchange through the TPN ligands, the Heisenberg spin Hamiltonian can be expressed as

$$H = J_1 \sum_{<i,j>_1} S_i \cdot S_j + J_2 \sum_{<i,j>_2} S_i \cdot S_j \tag{S1}$$

where $J_1$, $J_2$ and $S$ denote the exchange coupling constants and the unit spin vector aligned along the c-axis, respectively. We identify two distinct superexchange pathways in this lattice: the nearest-neighbor pathway along the square edge defined as $J_1$, corresponding to the 90° ligand–metal–ligand bridging geometry; and the next-nearest-neighbor diagonal pathway denoted as $J_2$, which arises from the 180° ligand–metal–ligand superexchange coupling. The spin quantum number for each ligand-based magnetic unit is set to $S = 1/2$.

By incorporating both the 90° $J_1$ edge-path and 180° $J_2$ diagonal-path ligand–metal–ligand superexchange interactions, the total energies of ferromagnetic (FM), altermagnetic (AM), and antiferromagnetic (AFM) spin configurations can be formulated as follows:

$$E_{AM} = E_0 - 8J_1S^2 + 4J_2S^2 \quad \text{(S2)}$$

$$E_{FM} = E_0 + 8J_1S^2 + 4J_2S^2 \quad (S3)$$

$$E_{AFM} = E_0 - 4J_2S^2 \quad (S4)$$

Accordingly, the exchange parameters $J_1$ and $J_2$ can be derived from the energy differences of distinct magnetic states:

$$J_1 = \frac{E_{FM} - E_{AM}}{16S^2} \quad \text{(S5)}$$

$$J_2 = \frac{8J_1S^2 - (E_{AFM} - E_{AM})}{8S^2} \quad \text{(S6)}$$

Herein, positive and negative signs correspond to parallel and antiparallel spin arrangements, respectively.

# Supporting Figures and Tables

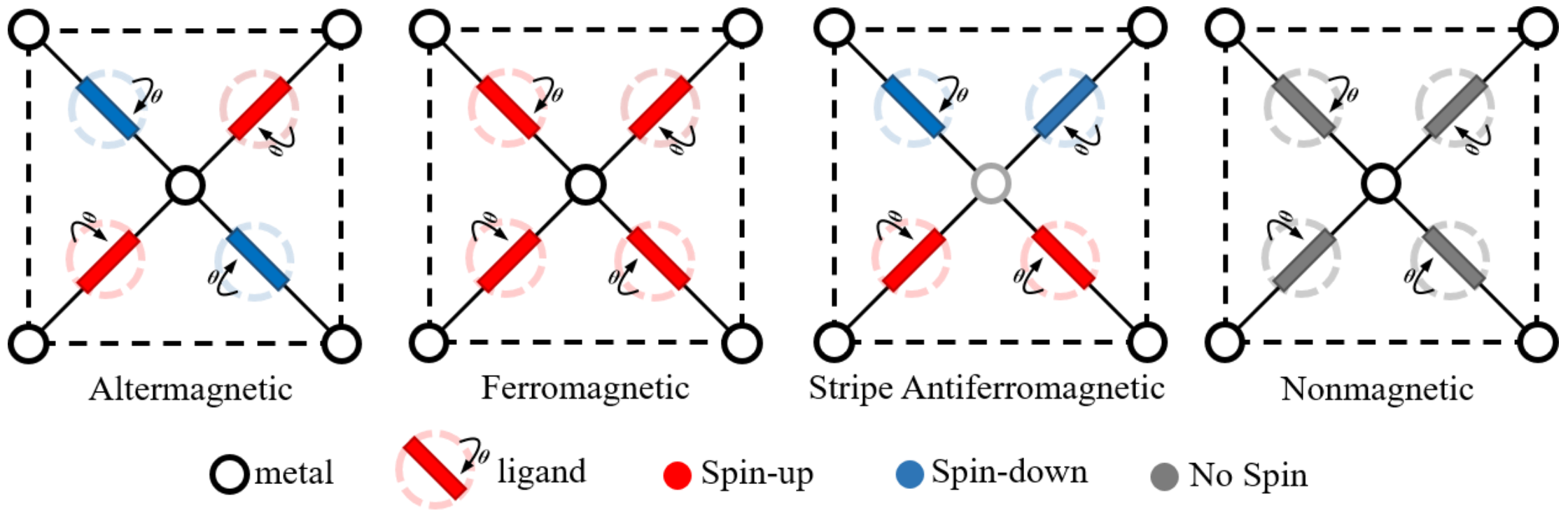

**Figure S1.** Four types of ligand-scale magnetic ordered structures modulated by ligand torsion angles (the dashed circles at the bottom correspond to the planar conformation at 0°, and the rectangular regions correspond to the twisted conformation at 90°): altermagnetic, ferromagnetic, stripe antiferromagnetic and nonmagnetic configurations. Red denotes spin-up ligands, blue denotes spin-down ligands, gray denotes spinless ligands, and hollow circles represent metal sites in the lattice.

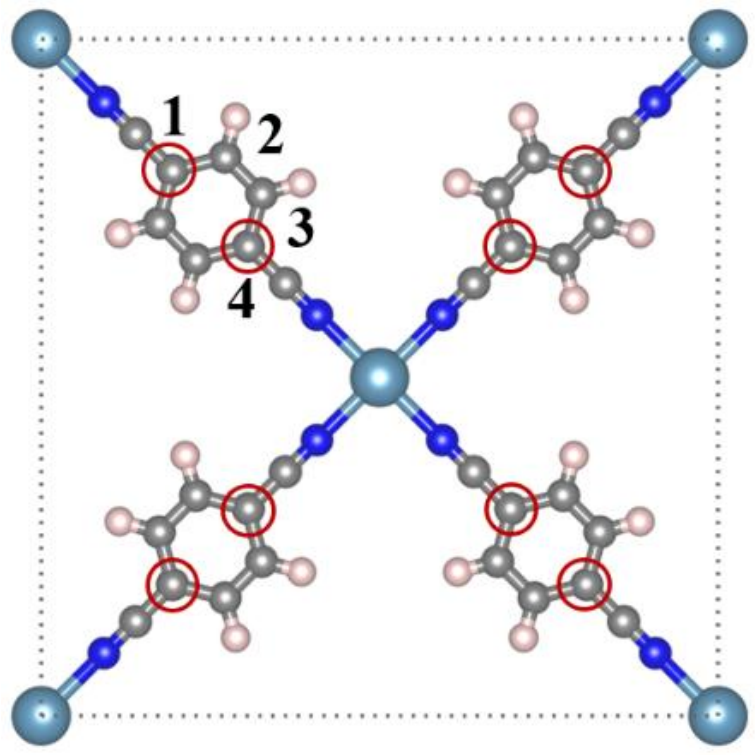

**Figure S2.** Selection of C and N Sites for PDOS Analysis. The eight circled red C atoms at the 1,4 positions and their eight corresponding N atoms are adopted to compute the partial density of states (PDOS) for the 0° and 90° conformations of pristine, fluorinated, and methylated $Ca(TPN)_2$. Teal, blue, gray, and white spheres represent Ca, N, C, and H atoms, respectively.

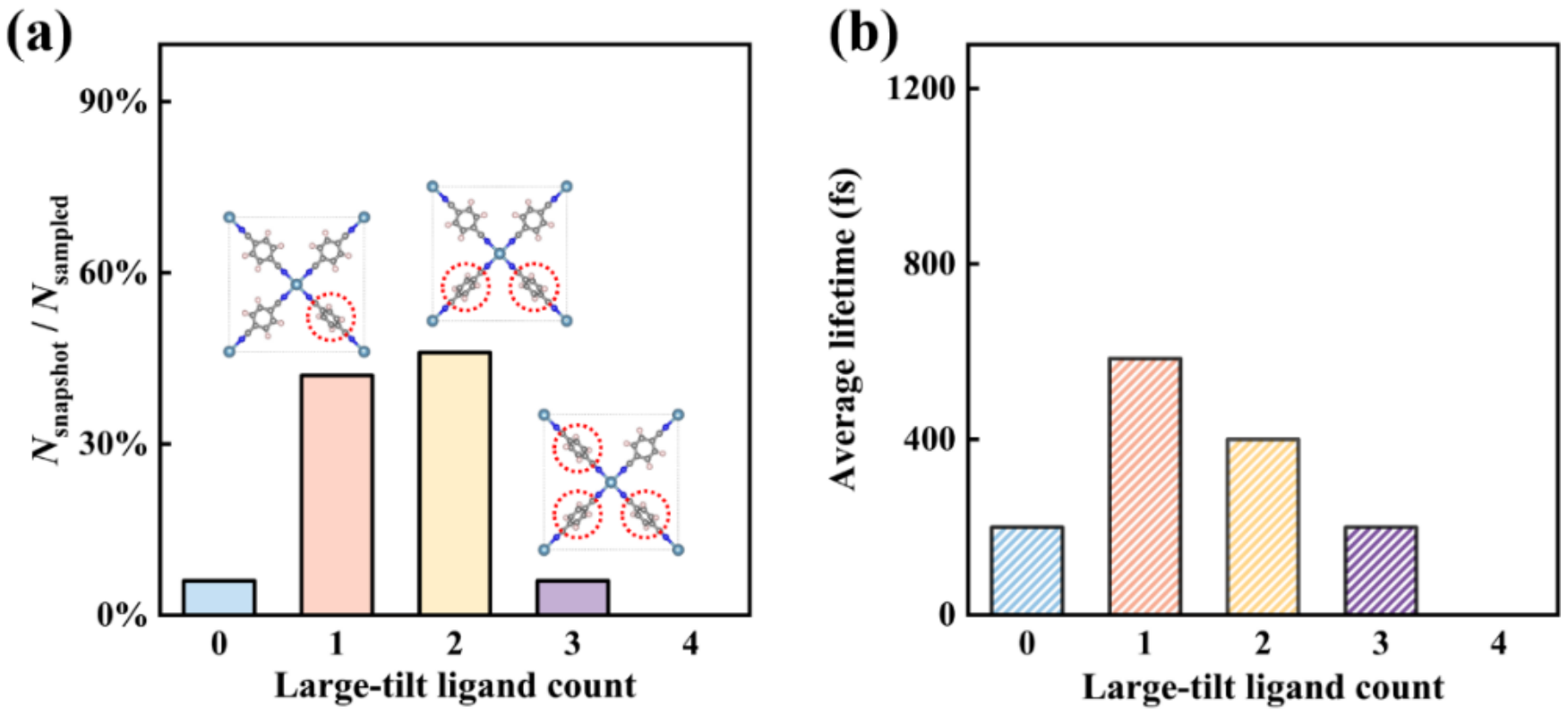

**Figure S3.** Ligand-flip statistics and lifetimes from AIMD simulations at 300 K. (a) Snapshot population classified by the number of ligands tilted beyond the critical angle (~34°), showing no-flip, single-ligand, dual-ligand, and three- or four-ligand flip events (bars), with representative structures of each event type shown in the insets. (b) Average lifetimes of the corresponding flipped configurations extracted from the trajectory.

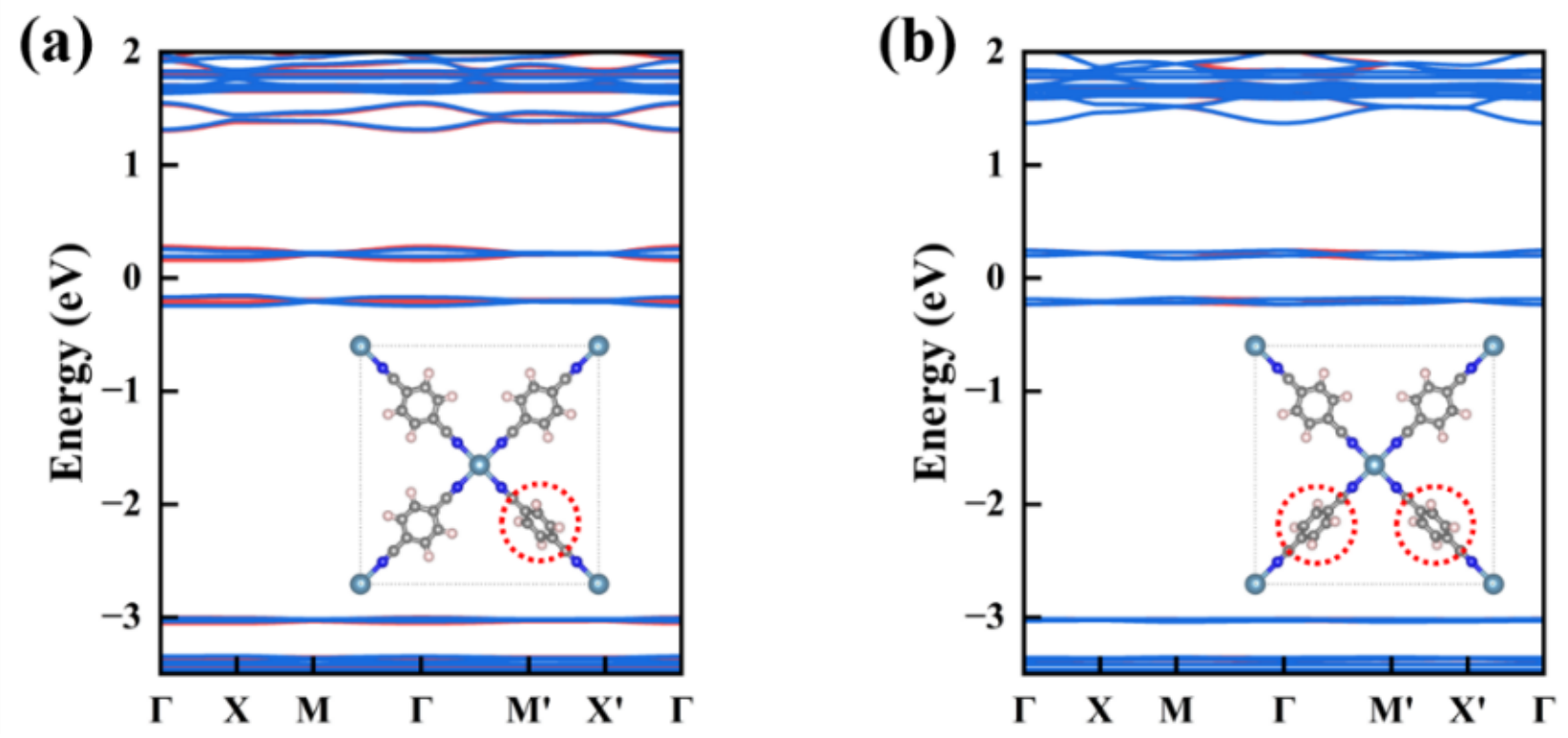

**Figure S4.** Band structures of thermally disordered ligand-flip snapshots: (a) Single flipped ligand. (b) Two flipped ligands. The spin splitting is strongly suppressed (~15 meV in (b)) relative to the fully ordered altermagnetic state.

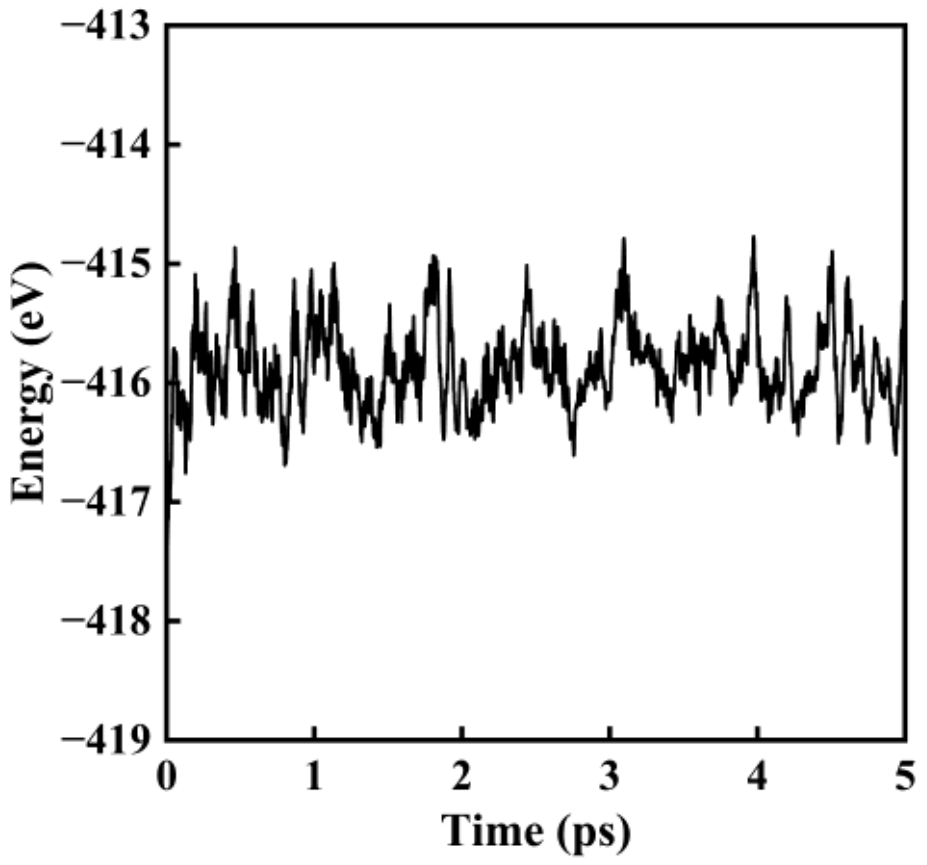

**Figure S5.** AIMD simulation of the fluorinated $Ca(TPN)_2$ monolayer at 300 K. Total-energy trajectory over 5 *ps*, confirming the structural integrity of the fluorinated framework at room temperature.

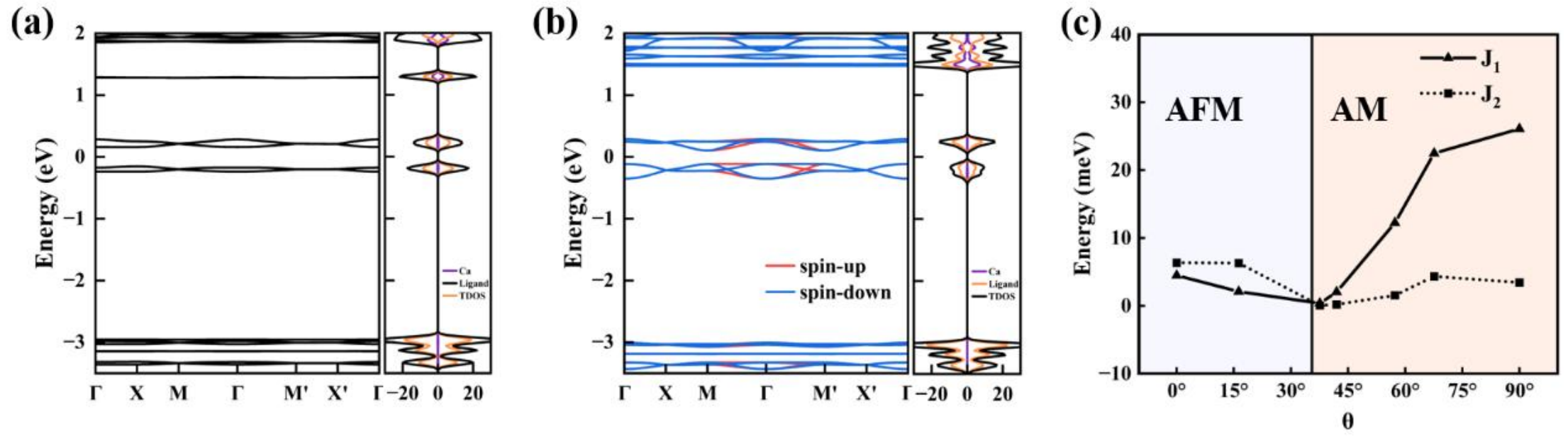

**Figure S6.** Electronic properties of the fluorinated $Ca(TPN)_2$ monolayer. (a) Band structure and PDOS of the 0° configuration in the AFM state. (b) Band structure and PDOS of the 90° configuration in the AM state. (c) Exchange parameters $J_1$ and $J_2$ as a function of ligand tilt angle.

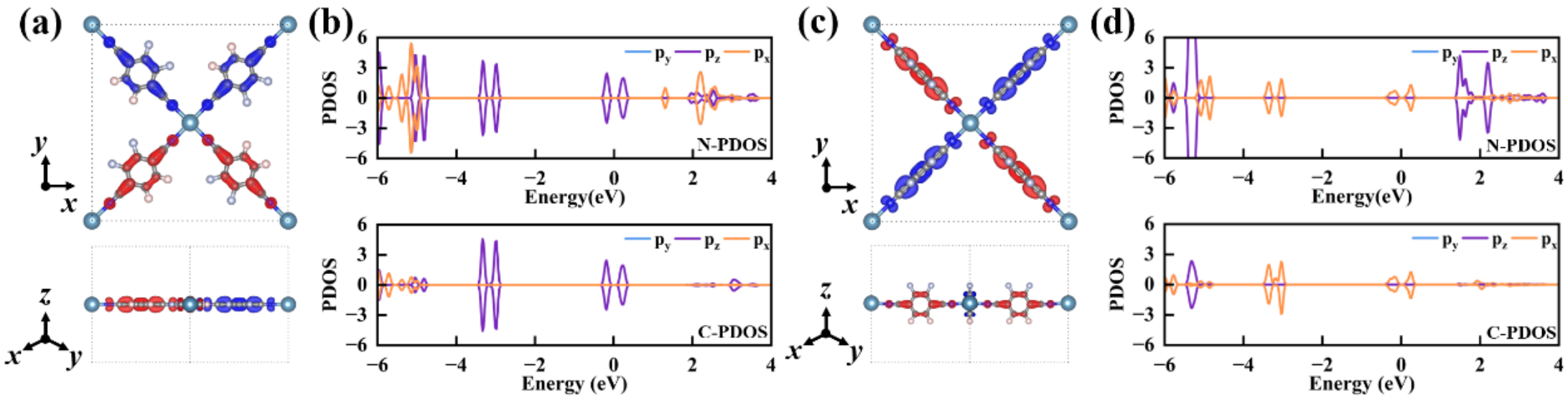

**Figure S7.** Spin charge density and projected density of states (PDOS) of the fluorinated $Ca(TPN)_2$ monolayer. (a,c) Top and side views of spin charge density for the 0° and 90° configurations, respectively. Red and blue isosurfaces denote spin-up and spin-down charge densities (isovalue = 0.004 e/bohr$^3$). (b,d) PDOS of the N and C atoms that contribute most to the magnetic moment in the 0° and 90° configurations, respectively.

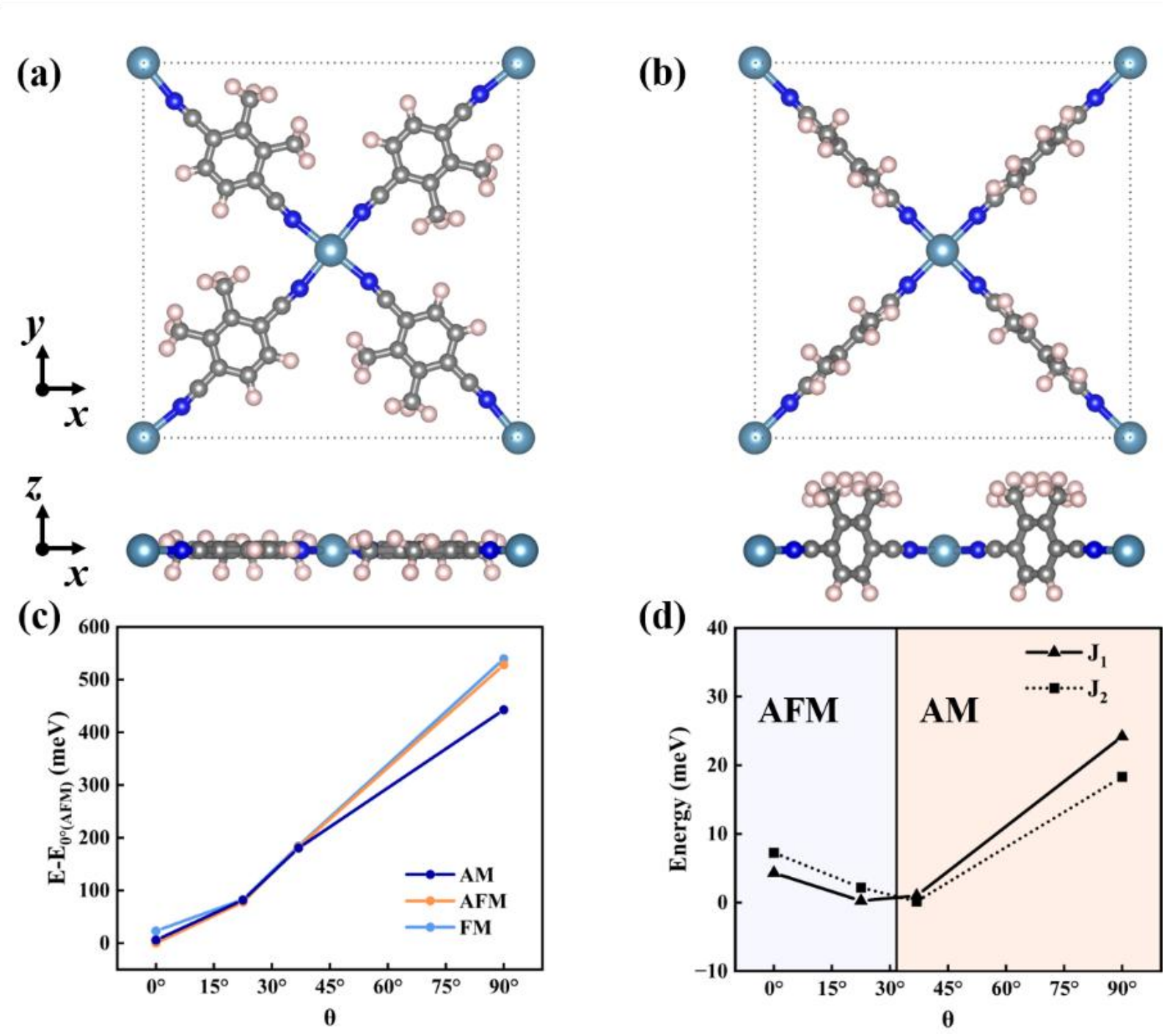

**Figure S8.** Structural, energetic, and magnetic exchange evolution of methylated Ca(TPN)$_2$ monolayer. (a) Planar configuration (0°). (b) Perpendicular configuration (90°), using a √2×√2 supercell. Top and side views are shown; teal, blue, gray, and white spheres represent Ca, N, C, and H atoms, respectively. (c) Relative energies of AFM, AM, and FM states as a function of ligand tilt angle, with the 0° AFM state set as zero. (d) Exchange parameters $J_1$ and $J_2$ as a function of ligand tilt angle.

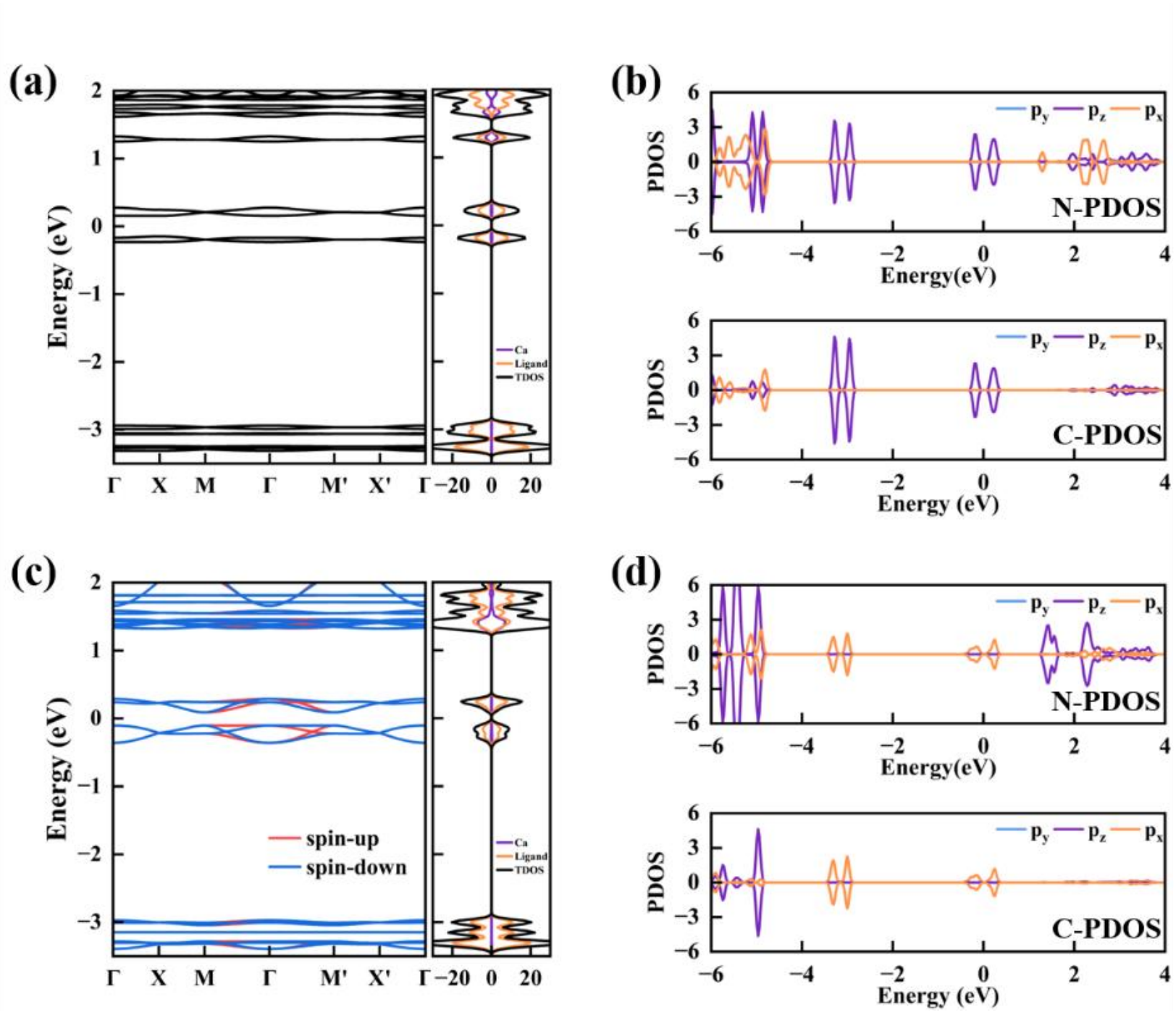

**Figure S9.** Electronic properties and projected density of states (PDOS) of methylated $Ca(TPN)_2$ monolayer. (a,c) Band structure and PDOS of the 0° and 90° configuration, respectively. (b,d) PDOS of the N and C atoms that contribute most to the magnetic moment in the 0° and 90° configurations, respectively.

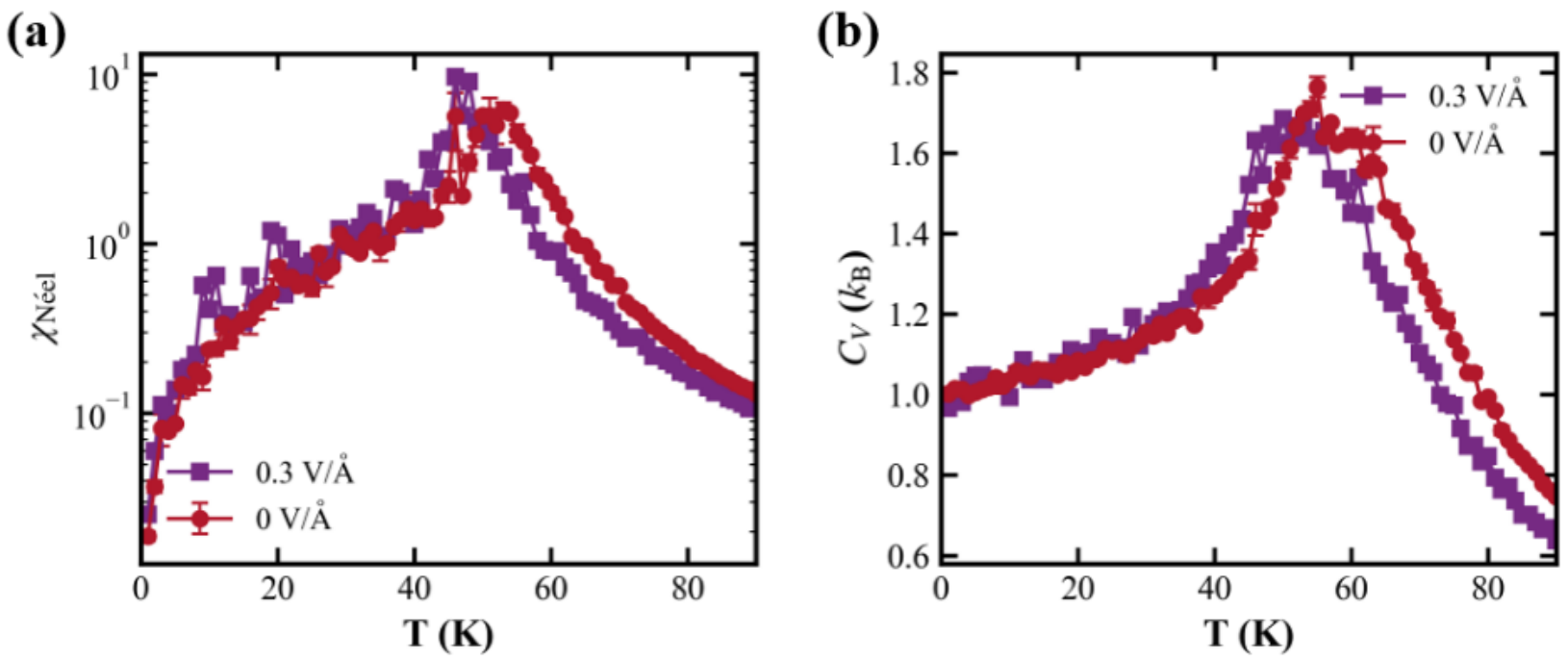

**Figure S10.** Finite-temperature magnetic response of the fluorinated 90° configuration under an applied electric field. (a) Staggered susceptibility $\chi_{\text{Néel}}$ and (b) specific heat $C_{\text{V}}$ (in units of $k_{\text{B}}$) as functions of temperature at zero field (red circles) and under 0.3 V/Å (purple squares). Both quantities shift to lower temperature under the field, corresponding to a reduction of the ordering scale from $T^*$ $\approx$54 K to $\approx$47 K.

**Table S1**. Lattice parameters ($L$ for length and $\alpha$ for angle), local magnetic moments ($\mu_M$) on metal cations and local magnetic moments ($\mu_L$) on ligand anions of $Ca(TPN)_2$ and fluorinated $Ca(TPN)_2$.

| | $Ca(TPN)_2(0°)$ | $Ca(TPN)_2(90°)$ | $Ca(hF\text{-}TPN)_2(0°)$ | $Ca(hF\text{-}TPN)_2(90°)$ |
|---|---|---|---|---|
| $L$ (Å) | 17.93 | 17.90 | 17.93 | 17.91 |
| $\alpha$ (°) | 90 | 90 | 90 | 90 |
| $\mu_M$ ($\mu_B$) | 0 | 0 | 0 | 0 |
| $\mu_L$ ($\mu_B$) | 0.50 | 0.43 | 0.50 | 0.43 |

* hF-TPN denotes the half-fluorinated TPN ligand.

**Table S2.** Energies (in meV per unit cell) of the altermagnetic (AM), ferromagnetic (FM), antiferromagnetic (AFM), and nonmagnetic (NM) states, and calculated nearest-neighbor ($J_1$) and next-nearest-neighbor ($J_2$) exchange parameters, for pristine and fluorinated $Ca(TPN)_2$ monolayers in the 0° and 90° configurations. All energies are referenced to the respective ground state.

| | $Ca(TPN)_2(0°)$ | $Ca(TPN)_2(90°)$ | $Ca(hF\text{-}TPN)_2(0°)$ | $Ca(hF\text{-}TPN)_2(90°)$ |
|---|---|---|---|---|
| $E_{AM}$ | 2.78 | 0 | 3.74 | 0 |
| $E_{FM}$ | 23.02 | 104.42 | 21.64 | 104.35 |
| $E_{AFM}$ | 0 | 60.07 | 0 | 58.97 |
| $E_{NM}$ | 245.02 | 103.53 | 246.36 | 103.56 |
| $J_1$ | 5.06 | 26.11 | 4.48 | 26.09 |
| $J_2$ | 6.45 | -3.94 | 6.35 | -3.40 |

[*] hF-TPN denotes the half-fluorinated TPN ligand.

**Table S3.** Calculated lattice parameters ($L$, in Å; $\alpha$, in °), local magnetic moments on metal cations ($\mu_M$) and ligand anions ($\mu_L$), magnetic energies ($E$, in meV per unit cell) referenced to the ground state, and exchange coupling constants $J_1$ and $J_2$ for methylated $Ca(TPN)_2$.

| | $L$ (Å) | $\alpha$ (°) | $\mu_M$ ($\mu_B$) | $\mu_L$ ($\mu_B$) | $E_{AM}$ | $E_{FM}$ | $E_{AFM}$ | $J_1$ | $J_2$ |
|---|---|---|---|---|---|---|---|---|---|
| 0° | 17.81 | 90 | 0 | 0.49 | 5.86 | 23.12 | 0 | 4.32 | 7.25 |
| 90° | 17.87 | 90 | 0 | 0.42 | 0 | 96.74 | 85.19 | 24.19 | -18.32 |

## References

(1) Kresse, G.; Furthmüller, J. Efficiency of ab-initio total energy calculations for metals and semiconductors using a plane-wave basis set. *Comput. Mater. Sci.* **1996,** *6*, 15-50.

(2) Kresse, G.; Furthmüller, J. Efficient iterative schemes for ab initio total-energy calculations using a plane-wave basis set. *Phys. Rev. B* **1996,** *54*, 11169-11186.

(3) Perdew, J. P.; Burke, K.; Ernzerhof, M. Generalized gradient approximation made simple. *Phys. Rev. Lett.* **1996,** *77*, 3865-3868.

(4) Grimme, S.; Ehrlich, S.; Goerigk, L. Effect of the damping function in dispersion corrected density functional theory. *J. Comput. Chem.* **2011,** *32*, 1456-65.

(5) Monkhorst, H. J.; Pack, J. D. Special points for Brillouin-zone integrations. *Phys. Rev. B* **1976,** *13*, 5188-5192.

(6) Nosé, S. A unified formulation of the constant temperature molecular dynamics methods. *J. Chem. Phys.* **1984,** *81*, 511-519.

(7) Hoover, W. G. Canonical dynamics: Equilibrium phase-space distributions. *Phys. Rev. A* **1985,** *31*, 1695-1697.

(8) Bylander, D. M.; Kleinman, L. Energy fluctuations induced by the Nose thermostat. *Phys. Rev. B* **1992,** *46*, 13756-13761.

(9) Nosé, S. Constant Temperature Molecular Dynamics Methods. *Prog. Theor. Phys. Suppl.* **1991,** *103*, 1-46.

(10) Wang, V.; Xu, N.; Liu, J.-C.; Tang, G.; Geng, W.-T. VASPKIT: A user-friendly interface facilitating high-throughput computing and analysis using VASP code. *Comput. Phys. Commun.* **2021,** *267*, 108033.